\documentclass[aip,jcp,reprint]{revtex4-1}
\usepackage{graphicx}
\usepackage{amsmath}
\usepackage{amssymb}

\begin{document}
\title{Electrolytes Between Dielectric Charged Surfaces: Simulations and Theory}

\author{Alexandre P. dos Santos}
\email{alexandre.pereira@ufrgs.br}
\affiliation{Instituto de F\'isica, Universidade Federal do Rio Grande do Sul, Caixa Postal 15051, CEP 91501-970, Porto Alegre, RS, Brazil.}

\author{Yan Levin}
\email{levin@if.ufrgs.br}
\affiliation{Instituto de F\'isica, Universidade Federal do Rio Grande do Sul, Caixa Postal 15051, CEP 91501-970, Porto Alegre, RS, Brazil.}

\begin{abstract}
We present a simulation method to study electrolyte solutions  in a dielectric slab geometry using a modified 3D Ewald summation. 
The method is fast and easy to implement, allowing us to rapidly resum an infinite series of image charges.
In the weak coupling limit, we also develop a mean-field theory which allows us to predict the ionic distribution between the dielectric charged plates. 
The agreement between both approaches, theoretical and simulational, is very good, validating both methods. Examples of ionic density profiles in the strong electrostatic coupling limit are also presented.  Finally, we explore the
confinement of charge asymmetric electrolytes between neutral surfaces.
\end{abstract}

\maketitle

\section{Introduction}\label{i}

Complex charged liquids present a variety of interesting phenomena such as like-charge attraction~\cite{KjMa86,LiLo99} and charge inversion~\cite{FeFe05,PiBa05,DoLe11}.  These phenomena arise as
a result of electrostatic correlations of counterions in the double layer~\cite{Le02}.  To explore the ionic
correlations between the overlapping double layers one must have accurate and reliable methods to study
electrolytes  in a slab geometry.  Similar difficulties are  encountered when one wants to understand the 
thermodynamics of ionic liquid based supercapacitors~\cite{ShKi12}.

Electrolytes in a slab-like geometry have been extensively studied in the past~\cite{RoBl96,Sh99,Ne01,MoNe02,SaTr11}. However, the dielectric contrast between the solvent and the confining surfaces is usually not taken into account. 
This dielectric mismatch leads to polarization of  interfaces 
which dramatically increases  the mathematical complexity of the problem. For a single interface, the dielectric heterogeneity and the resulting 
induced surface charge can lead to a repulsion of electrolyte from the charged dielectric surface~\cite{MoNe00,PoDa14}. Similar behavior has been observed near charged colloidal particles~\cite{Me02,DoBa11,BaDo11,LuLi11,DiDo12,JaSo13,GaDe14}.

Most of the theoretical and simulational works involving confined electrolytes neglect the dielectic contrast which results in an infinite series of image charges.  While this significantly simplifies the calculations, it also fails to account for some of the fundamental physics of the overlapping  double layers. Recently Wang and Wang~\cite{WaWa13} presented a mean-field theoretical discussion of confined electrolytes between charged and neutral plates. In the same year, Zwanikken and de la Cruz~\cite{ZwDe13} developed a liquid state theory which predicted that neutral confining polarizable interfaces can attract each other
inside an electrolyte solution. Similar result can be found in other works~\cite{BeLe68,KjMa87,Kana12}. Samaj and Trizac~\cite{SaTr12} and Jho~{\it et~al.}~\cite{JhKa08} developed theories to study the distribution  of confined counterions between charged plates, in a salt free system in a strong-coupling limit. Some studies focused
on the specific case in which the dielectric constant of the surrounding medium is much lower than of water~\cite{HaHa89,PeRa96,Kl06}.
Jho~{\it et~al.}~\cite{JhPa07} developed a simulation method for confined counterions based on the electrostatic layer correction~(ELC) method~\cite{ArJo02}. Also, Tyagi~{\it et~al.}~\cite{TyAr07,TyAr08} constructed the ICMMM2D method which is the generalization of the MMM2D algorithm~\cite{ArHo02}, previously developed to study homogeneous dielectric slab systems. Although these methods account for the surface polarization, they require a calculation of a sum of terms for the electrostatic potential that make  simulations quite slow. Similar difficulties are encountered with other simulation approaches~\cite{Kana12}. Boundary element methods (BEM) consider the minimization of functionals and can be applied to systems with general geometries. Some BEM methods consider local polarization charge densities as dynamic variables~\cite{MaBo01,JaSo12} others, attempt to explicitly calculate the bound charge~\cite{BoGi04,BaSi14}. Even though BEM methods are expensive computationally, they have been extensively used to study general soft matter problems~\cite{NaHe11,BoHe13,BeFu14}.

In this paper we present a simulation method based on 3D Ewald summation with a modified  correction for the slab geometry~\cite{YeBe99}. The method is  simple to implement. It does not require summations of  a slowly convergent infinite series of images  during the simulation, and is comparable in time with a regular 3D Ewald method. The paper is organized as follow. In Section~\ref{ee}, we show how to construct the electrostatic energy of the system.  In Section~\ref{mcs}, we study confined electrolytes between polarizable charged surfaces. In Section~\ref{mpbe}, we present a mean-field theory for 1:1 electrolytes
and compare it with simulations. 
In Section~\ref{r}, we present the general results, and in the Section~\ref{c}, the conclusions.

\section{Electrostatic Energy}\label{ee}

\begin{figure}[h]
\begin{center}
\includegraphics[width=4.5cm]{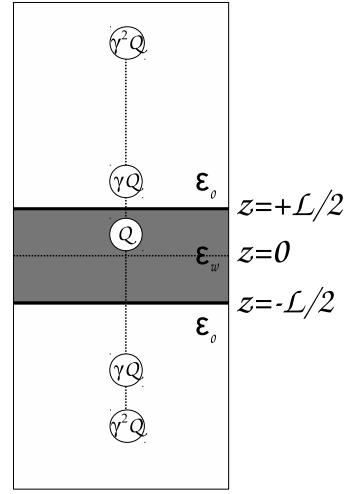}
\end{center}
\caption{Representation of a charge between the dielectric surfaces. Only the first and second order images are shown,  $N_i=2$.}
\label{fig1}
\end{figure}

To perform simulations, we use a rectangular simulation box of sides $L_x$, $L_y$ and $L_z$. 
The box contains  $N$ ions of charges $q_j=\alpha_j Q$, where $\alpha_j$ is the valence of the ion
and $Q$ is the proton charge, confined in the region $-L_x/2<x<L_x/2$, $-L_y/2<y<L_y/2$ and $-L/2<z<L/2$. We set $L_y=L_x$ and $L_z=5 L$. The uniform dielectric constants are: $\epsilon_w$ inside the slab containing electrolyte, and $\epsilon_o$ outside. 
The dielectric contrast results in an infinite set of ``images of images" which  must be resumed to obtain the total electrostatic energy.  We define  $N_i$ as a number of images of an ion at each interface.  To calculate the exact electrostatic energy, 
$N_i$ should be infinite.  This, however, is not practical in a simulation.  Instead, we explore 
the convergence of simulations as the number of images $N_i$ is increased.  For example, if $N_i=2$, we consider one image charge at each dielectric interface and the image of image, producing 4 image charges for each ion, see Fig.~\ref{fig1}. The electrostatic potential at the position ${\pmb r}$ (in the region with $\epsilon_w$), created by all ions (excluding ion $i$), their image charges (including the image of ion $i$), and the periodic replicas is
\begin{eqnarray}\label{elec_pot}
\phi_i({\pmb r})=\sum_{\pmb n}^{\infty} \sum_{j=1}^{N}{}^{'}\int\frac{\rho_j({\pmb s})}{\epsilon_w
|{\pmb r}-{\pmb s}|}d^3{\pmb s} + \nonumber \\
\sum_{m=1}^{N_i} \sum_{\pmb n}^{\infty} \sum_{j=1}^{N} \left[ \int\frac{\rho_{j m}^{+}({\pmb s})}{\epsilon_w
|{\pmb r}-{\pmb s}|}d^3{\pmb s}  \right. + \nonumber \\
\left. \int\frac{\rho_{j m}^{-}({\pmb s})}{\epsilon_w
|{\pmb r}-{\pmb s}|}d^3{\pmb s} \right] \ ,
\end{eqnarray}
where $\rho_j({\pmb s})=q_j \delta({\pmb s}-{\pmb r}_j-{\pmb r}_{ep})$ and
$\rho_{j m}^{\pm}({\pmb s})=\gamma^m q_j \delta({\pmb s}-{\pmb r}^{\pm}_{j m}-{\pmb r}_{ep})$ are the
charge densities of ions and their replicas  and of dielectric images
and their replicas.    

The replication vector is defined as ${\pmb
r}_{ep}=L_x n_x\hat{\pmb x}+L_y n_y\hat{\pmb y}+L_z n_z\hat{\pmb z}$ and ${\pmb r}_{j m}^{\pm}=x_j \hat{\pmb x} + y_j \hat{\pmb y} + \left[ (-1)^m z_j \pm m L \right] \hat{\pmb z}$. The
vectors ${\pmb n}=(n_x,n_y,n_z)$, where $n_x$, $n_y$ and $n_z$ are positive and negative integers, 
represent the infinite replicas  of the main
cell. The constant $\gamma$ is
defined as $\gamma=(\epsilon_w-\epsilon_o)/(\epsilon_w+\epsilon_o)$ and the
prime on the summation signifies that $j\neq i$, when ${\pmb n}=(0,0,0)$. The total
electrostatic energy of the system is
\begin{equation}
U=\frac{1}{2}\sum_{i=1}^{N}q_i\phi_i({\pmb r}_i) \ .
\end{equation} 

The energy above is very difficult to calculate because of a slow conditional 
convergence  of the series in Eq.~\eqref{elec_pot}. To speed up the convergence, 
we use the Ewald method in which the ionic charge is 
partially screened by placing a Gaussian-distributed
charge of opposite sign on top of each ion~\cite{AlTi87}. We then add and subtract an opposite Gaussian charge at the position of each ion 
and its images, $\rho_j({\pmb s})$ and $\rho_{j m}^{\pm}({\pmb s})$, respectively.
The potential, Eq.~\eqref{elec_pot}, then becomes
\begin{equation}\label{elec_pot2}
\phi_i({\pmb r})=\phi_i^S({\pmb r})+\phi^L({\pmb r})-\phi_i^{self}({\pmb r}) \ ,
\end{equation}
where
\begin{eqnarray}\label{phi_S}
\phi_i^S({\pmb r})=\sum_{\pmb n}^{\infty}\sum_{j=1}^{N}{}^{'}\int\frac{\rho_j({\pmb s})-\rho_j^G({\pmb
s})}{\epsilon_w |{\pmb r}-{\pmb s}|}d^3{\pmb s} + \nonumber \\
\sum_{m=1}^{N_i} \sum_{\pmb n}^{\infty} \sum_{j=1}^{N} \left[ \int\frac{\rho_{j m}^{+}({\pmb s})-\rho_{j m}^{G+}({\pmb s})}{\epsilon_w
|{\pmb r}-{\pmb s}|}d^3{\pmb s}  \right. + \nonumber \\
\left. \int\frac{\rho_{j m}^{-}({\pmb s})-\rho_{j m}^{G-}({\pmb s})}{\epsilon_w
|{\pmb r}-{\pmb s}|}d^3{\pmb s} \right] \ ,
\end{eqnarray}
\begin{eqnarray}\label{phi_L}
\phi^L({\pmb r})=\sum_{\pmb n}^{\infty}\sum_{j=1}^{N}\int\frac{\rho_j^G({\pmb s})}{\epsilon_w |{\pmb
r}-{\pmb s}|}d^3{\pmb s} + \nonumber \\
\sum_{m=1}^{N_i} \sum_{\pmb n}^{\infty} \sum_{j=1}^{N} \left[ \int\frac{\rho_{j m}^{G+}({\pmb s})}{\epsilon_w
|{\pmb r}-{\pmb s}|}d^3{\pmb s}  \right. + \nonumber \\
\left. \int\frac{\rho_{j m}^{G-}({\pmb s})}{\epsilon_w
|{\pmb r}-{\pmb s}|}d^3{\pmb s} \right]
\end{eqnarray}
and
\begin{equation}\label{phi_self}
\phi^{self}_i({\pmb r})= \int\frac{\rho_i^G({\pmb s})}{\epsilon_w |{\pmb r}-{\pmb
s}|}d^3{\pmb s} \ ,
\end{equation}
where
\begin{equation}
\rho_j^G({\pmb
s})=q_j (\kappa_e^3/\sqrt{\pi^3})\exp{(-\kappa_e^2|{\pmb s}-{\pmb r}_j-{\pmb
r}_{ep}|^2)} \ ,
\end{equation}
\begin{equation}
\rho_{jm}^{G\pm}({\pmb s})=\gamma^m
q_j(\kappa_e^3/\sqrt{\pi^3})\exp{(-\kappa_e^2|{\pmb s}-{\pmb r}_{j m}^{\pm}-{\pmb
r}_{ep}|^2)} \ ,
\end{equation}
and $\kappa_e$ is a damping parameter. Note that we have subtracted the self
potential, Eq.~\eqref{phi_self}, from Eq.~\eqref{elec_pot2}, in order
to remove the prime over the summation in the long-range ($L$) 
part of the potential, Eq.~\eqref{phi_L}. The electrostatic potential produced by the Gaussian charges
can be calculated using the Poisson equation, yielding
\begin{eqnarray}\label{phi_L2}
\phi^L({\pmb r})=\sum_{\pmb n}^{\infty}\sum_{j=1}^{N}
q_j\frac{\text{erf}{(\kappa_e |{\pmb r}-{\pmb r}_j-{\pmb r}_{ep}|)}}{\epsilon_w
|{\pmb r}-{\pmb r}_j-{\pmb r}_{ep}|} + \nonumber \\
\sum_{m=1}^{N_i} \sum_{\pmb n}^{\infty} \sum_{j=1}^{N}\gamma^m q_j \left[ \frac{\text{erf}{(\kappa_e
|{\pmb r}-{\pmb r}^+_{jm}-{\pmb r}_{ep}|)}}{\epsilon_w |{\pmb r}-{\pmb r}^+_{jm}-{\pmb r}_{ep}|} \right. + \nonumber \\
\left. \frac{\text{erf}{(\kappa_e|{\pmb r}-{\pmb r}^-_{jm}-{\pmb r}_{ep}|)}}{\epsilon_w |{\pmb r}-{\pmb r}^-_{jm}-{\pmb r}_{ep}|} \right] \ ,
\end{eqnarray}
where $\text{erf}(x)$ is the error function. The short-range part of the 
potential ($S$), Eq.~\eqref{phi_S}, can then be obtained 
in terms of the complementary error function,
$\text{erfc}(x)=1-\text{erf}(x)$,
\begin{eqnarray}\label{phi_S2}
\phi_i^S({\pmb r})=\sum_{\pmb n}^{\infty}\sum_{j=1}^{N}{}^{'}
q_j\frac{\text{erfc}{(\kappa_e |{\pmb r}-{\pmb r}_j-{\pmb r}_{ep}|)}}{\epsilon_w
|{\pmb r}-{\pmb r}_j-{\pmb r}_{ep}|} + \nonumber \\
\sum_{m=1}^{N_i} \sum_{\pmb n}^{\infty} \sum_{j=1}^{N}\gamma^m q_j \left[ \frac{\text{erfc}{(\kappa_e
|{\pmb r}-{\pmb r}^+_{jm}-{\pmb r}_{ep}|)}}{\epsilon_w |{\pmb r}-{\pmb r}^+_{jm}-{\pmb r}_{ep}|} \right. + \nonumber \\
\left. \frac{\text{erfc}{(\kappa_e|{\pmb r}-{\pmb r}^-_{jm}-{\pmb r}_{ep}|)}}{\epsilon_w |{\pmb r}-{\pmb r}^-_{jm}-{\pmb r}_{ep}|} \right] \ .
\end{eqnarray}
This potential decays exponentially fast, with the decay length controlled by the damping parameter which we set to $\kappa_e=4/R_c$, where $R_c=L_x$ is the distance cutoff. It is then necessary to consider only the term ${\pmb n}=(0,0,0)$,  with the usual periodic boundary condition.  Furthermore, for sufficiently 
large values of $\kappa$ its is sufficient to include only a few
images-of-images. The damping parameter, however, can not be too high since its value controls the number of ${\pmb k}$-vectors that will have to be used to calculate the long-range potential. For systems studied in this paper, we find that  $N_i=2$, {\it in the short range potential}, is sufficient. Depending on the separation between the plates, more images may be necessary. Prior to accumulation of data we, therefore,  check for convergence by varying the value of $N_i$. The short-range potential then becomes
\begin{eqnarray}\label{phi_S3}
\phi_i^S({\pmb r})=\sum_{j=1}^{N}{}^{'}
q_j\frac{\text{erfc}{(\kappa_e |{\pmb r}-{\pmb r}_j|)}}{\epsilon_w
|{\pmb r}-{\pmb r}_j|} + \nonumber \\
\sum_{m=1}^{2} \sum_{j=1}^{N}\gamma^m q_j \left[ \frac{\text{erfc}{(\kappa_e
|{\pmb r}-{\pmb r}^+_{jm}|)}}{\epsilon_w |{\pmb r}-{\pmb r}^+_{jm}|} \right. + \nonumber \\
\left. \frac{\text{erfc}{(\kappa_e|{\pmb r}-{\pmb r}^-_{jm}|)}}{\epsilon_w |{\pmb r}-{\pmb r}^-_{jm}|} \right] \ .
\end{eqnarray}
The self-potential, Eq.~\eqref{phi_self}, reduces to
\begin{equation}\label{phi_self2}
\phi_i^{self}({\pmb r})=q_i\frac{\text{erf}{(\kappa_e |{\pmb r}-{\pmb
r}_i|)}}{\epsilon_w |{\pmb r}-{\pmb r}_i|} \ .
\end{equation}
We next calculate the long-range part of the potential, Eq.~\eqref{phi_L2}.
This is most easily obtained using the Fourier representation,  $\hat{\phi}^L({\pmb
k})=\int d^3{\pmb r}\ \exp{(-i{\pmb k}\cdot{\pmb r})}\phi^L({\pmb
r})$, since in the reciprocal space all sums, once again, converge very rapidly. The Fourier transform $\hat{\rho}^T({\pmb
k})=\int d^3{\pmb r}\ \exp{(-i{\pmb k}\cdot{\pmb r})}\rho^T({\pmb r})$, of the Gaussian charge density,
\begin{eqnarray}
\rho^T({\pmb r})=\sum_{\pmb n}^{\infty}\sum_{j=1}^{N} q_j\frac{\kappa_e^3}{\sqrt{\pi^3}}
\exp{(-\kappa_e^2|{\pmb r}-{\pmb r}_j-{\pmb r}_{ep}|^2)} + \nonumber \\
\sum_{m=1}^{N_i}\sum_{\pmb n}^{\infty}\sum_{j=1}^{N}\gamma^m q_j\frac{\kappa_e^3}{\sqrt{\pi^3}} \left[ \exp{(-\kappa_e^2|{\pmb r}-{\pmb r}^+_{jm}-{\pmb r}_{ep}|^2)} \right. + \nonumber \\
\left. \exp{(-\kappa_e^2|{\pmb r}-{\pmb r}^-_{jm}-{\pmb r}_{ep}|^2)} \right] \ ,
\end{eqnarray}
is 
\begin{eqnarray}
\hat{\rho}^T({\pmb k})= \frac{(2\pi)^3}{V} \exp{(-\frac{|{\pmb
k}|^2}{4\kappa_e^2})}\left[\sum_{j=1}^{N}q_j\exp{(-i{\pmb k}\cdot{\pmb r}_j)} \right.+ \nonumber \\
\left. \sum_{m=1}^{N_i}\sum_{j=1}^{N}\gamma^m q_j \left[ \exp{(-i{\pmb k}\cdot{\pmb r}^+_{jm})}+\exp{(-i{\pmb k}\cdot{\pmb r}^-_{jm})} \right] \right] \ ,
\end{eqnarray}
where ${\pmb k}=(2\pi n_x/L_{xy},2\pi n_y/L_{xy},2\pi n_z/L_{z})$ and $V=L_x L_y L_z$. Using
the Poisson equation, $|{\pmb k}|^2 \hat{\phi}^L({\pmb
k})=(4\pi/\epsilon_w) \hat{\rho}^T({\pmb k})$, we can evaluate the Fourier
transform of the potential,
\begin{eqnarray}
\hat{\phi}^L({\pmb k})=\frac{8\pi^4}{\epsilon_w V |{\pmb k}|^2}\exp{(-\frac{|{\pmb k}|^2}{4\kappa_e^2})} \left[ \sum_{j=1}^{N}q_j\exp{(-i{\pmb k}\cdot{\pmb r}_j)} \right.+\nonumber \\
\left. \sum_{m=1}^{N_i}\sum_{j=1}^{N}\gamma^m q_j \left[ \exp{(-i{\pmb k}\cdot{\pmb r}^+_{jm})}+ \exp{(-i{\pmb k}\cdot{\pmb r}^-_{jm})} \right] \right] \ .
\end{eqnarray}
The corresponding real-space  electrostatic potential is calculated using the inverse Fourier
transform, $\phi^L({\pmb r})=\dfrac{1}{(2\pi)^3}\sum_{{\pmb k}}\hat{\phi}^L({\pmb k})\exp{(i{\pmb k}\cdot{\pmb r})}$,
\begin{eqnarray}\label{phi_L3}
 \phi^L({\pmb r})=\sum_{{\pmb k}}\frac{4\pi}{\epsilon_w V |{\pmb k}|^2}
\exp{(-\frac{|{\pmb k}|^2}{4\kappa_e^2})}\exp{(i{\pmb k}\cdot{\pmb r})}\times \nonumber \\
\left[\sum_{j=1}^{N}q_j\exp{(-i{\pmb k}\cdot{\pmb r}_j)} \right.+ \nonumber \\
\left. \sum_{m=1}^{N_i}\sum_{j=1}^{N}\gamma^m q_j \left[ \exp{(-i{\pmb k}\cdot{\pmb r}^+_{jm})} + \exp{(-i{\pmb k}\cdot{\pmb r}^-_{jm})} \right] \right] \ .
\end{eqnarray}

The long-range contribution to the total electrostatic energy is then given by
$U_L=(1/2)\sum_{i=1}^{N}q_i\phi^L({\pmb r}_i)$, where $\phi^L({\pmb
r})$ is obtained from Eq.~\eqref{phi_L3}. It is convenient to rewrite this 
in terms of functions:

\begin{equation}
A({\pmb k})= \sum_{i=1}^{N}q_i \cos{({\pmb k}\cdot {\pmb
r}_i)} \ ,
\end{equation}

\begin{equation}
B({\pmb k})=- \sum_{i=1}^{N}q_i \sin{({\pmb k}\cdot {\pmb r}_i)} \ ,
\end{equation}

\begin{equation}\label{sum1}
C({\pmb k})= \sum_{m=1}^{N_i} \sum_{i=1}^{N} \gamma^m q_i \left[ \cos{({\pmb k}\cdot {\pmb r}^+_{im})} + \cos{({\pmb k}\cdot {\pmb r}^-_{im})} \right]
\end{equation}

and

\begin{equation}\label{sum2}
D({\pmb k})=- \sum_{m=1}^{N_i} \sum_{i=1}^{N} \gamma^m q_i \left[ \sin{({\pmb k}\cdot {\pmb r}^+_{im})} + \sin{({\pmb k}\cdot {\pmb r}^-_{im})} \right] \ .
\end{equation}

The electrostatic energy then becomes,
\begin{eqnarray}\label{U_long}
U_L = \sum_{{\pmb k}}\frac{2\pi}{\epsilon_w V |{\pmb k}|^2}
\text{exp}(-\frac{|{\pmb k}|^2}{4\kappa_e^2}) \times \nonumber \\
\left[A({\pmb k})^2 + B({\pmb k})^2 + A({\pmb k}) C({\pmb k}) + B({\pmb k}) D({\pmb k})\right] \ .
\end{eqnarray}
The terms in Eqs.~\eqref{sum1} and \eqref{sum2}, are multiplied by the $\gamma^m$ parameter, leading to a converging sum for realistic $\gamma<1$ parameter values. However, we do not know {\it a priori} a minimum number of images necessary to obtain an accurate result for the long-range potential.  For example,  we find that for $\gamma\approx 0.9$,  we need $N_i=50$ to obtain a good convergence. However, such a large number of images  makes simulations extremely slow.  We note, however, that  Eqs.~\eqref{sum1} and \eqref{sum2} can be rewritten as,
\begin{eqnarray}
C({\pmb k}) = \sum_{i=1}^{N} q_i \left[ c_1({\pmb k}) \cos{(k_x x_i+k_y y_i)} \cos{(k_z z_i)}\right. +\nonumber \\
\left. c_2({\pmb k}) \sin{(k_x x_i+k_y y_i)} \sin{(k_z z_i)}\right]
\end{eqnarray}
and
\begin{eqnarray}
D({\pmb k})= - \sum_{i=1}^{N} q_i \left[ d_1({\pmb k}) \sin{(k_x x_i+k_y y_i-k_z z_i)}\right. +\nonumber \\
\left. d_2({\pmb k}) \sin{(k_x x_i+k_y y_i+k_z z_i)}\right] \ .
\end{eqnarray}

where,
\begin{equation*}
c_1({\pmb k})=2\sum_{m=1}^{N_i}\gamma^m \cos{(m k_z L)} \ ,
\end{equation*}
\begin{equation*}
c_2({\pmb k})=2\sum_{m=1}^{N_i}(-1)^{m+1}\gamma^m \cos{(m k_z L)} \ ,
\end{equation*}
\begin{equation}
d_1({\pmb k})=2\sum_{m_o=1}^{N_i}\gamma^{m_o} \cos{(m_o k_z L)} \ ,
\end{equation}
\begin{equation}
d_2({\pmb k})=2\sum_{m_e=2}^{N_i} \gamma^{m_e} \cos{(m_e k_z L)} \ ,
\end{equation}
and $m$ are integers: $m_o$ are odd and $m_e$ are even.  The parameters $c_1({\pmb k})$, $c_2({\pmb k})$, $d_1({\pmb k})$ and $d_2({\pmb k})$ can be obtained once (up to any desired accuracy) at the beginning of the simulation, since they do not depend on the ionic positions. 
The functions, $A({\pmb k})$, $B({\pmb k})$, $C({\pmb k})$ and $D({\pmb k})$, can  now be easily updated for each new configuration in a Monte Carlo~(MC) simulation.

The electrostatic energy resulting from the short-range part of the potential is $U_S=(1/2)\sum_{i=1}^{N}q_i\phi_i^S({\pmb r}_i)$, where $\phi_i^S({\pmb r})$ is given by the Eq.~\eqref{phi_S3}, and the self-energy contribution is $U_{self}=(1/2)\sum_{i=1}^{N}q_i\phi_i^{self}({\pmb r}_i)$. In the limit $x\rightarrow 0$, the $\text{erf}(x)$ function vanishes as $(2/\sqrt{\pi})x$ and the self-energy contribution reduces to $U_{self}=(\kappa_e/\epsilon_w\sqrt{\pi})\sum_{i=1}^{N}q_i^2$.
The total electrostatic interaction energy of the ions is given by the expressions above, plus the correction needed to account for the slab geometry~\cite{YeBe99}.  

Yeh and Berkowitz~\cite{YeBe99} found that a regular 3D Ewald summation method with an energy correction which accounts for the anisotropic summation of a conditionally convergent series in a slab-like geometry can reproduce the same results as the 2D Ewald method, with a significant gain in performance. For more details on the Ewald summation method, applied to different geometries, an interested reader can consult Refs.~\cite{Sp97,Sm08,Ba14}.
Following Yeh and Berkowitz and Smith~\cite{Sm81} and taking into account the dielectric discontinuity and the induced image charges,
we find the correction for the slab geometry to be
\begin{eqnarray}
U_{cor}=-\frac{\pi}{\epsilon_w V}\sum_{i=1}^{N}q_i \left[ \sum_{j=1}^{N} q_j
(z_i-z_j)^2 + \right.\nonumber \\
\left.\sum_{m=1}^{N_i} \sum_{j=1}^{N} \gamma^m q_j \left[ (z_i-z^-_{jm})^2 + (z_i-z^+_{jm})^2 \right] \right] \ ,
\end{eqnarray}
where $z^{\pm}_{jm}=(-1)^m z_j \pm m L$. Using the electroneutrality, this
expression can be rewritten as
\begin{equation}\label{U_cor}
U_{cor}=\frac{2\pi}{\epsilon_w V} M_z^2 \left[ 1+2\sum_{m=1}^{N_i}(-\gamma)^m\right] \ ,
\end{equation}
where $M_z=\sum_{i=1}^{N}q_i z_i$ is the magnetization in the $\hat{z}$ direction. Again, the constant between the brackets can be evaluated once at the beginning of the simulation, so that  $M_z$ can be easily updated in the simulation process. The total energy of the system is then 
\begin{equation}
U=U_S+U_L+U_{cor}+U_{self} \ .
\end{equation}

\section{Monte Carlo simulations}\label{mcs}

We now study an electrolyte solution confined between two negatively charged dielectric surfaces. 
The two charged plates contain $256$ point charges each, uniformly distributed on the surface on a square lattice.
The magnitude of point charges is adjusted to obtain the desired surface charge density.  The surfaces are
located at $z=-L/2$ and $z=+L/2$.  Counterions are modeled as hard spheres with the charge $q$ located at the center. Besides the counterions, salt ions can also be present in the system, all with the same ionic radius, $2$~\AA. The solvent is modeled as an uniform dielectric medium. The Bjerrum length $\lambda_B=\beta Q^2/\epsilon_w$ of the system is $7.2~$\AA, corresponding  to  water at room temperature. The MC simulations are performed using the Metropolis algorithm. The method developed in Section~\ref{ee} is used to obtain the electrostatic energy. Care must be taken in the calculation of the electrostatic energy of wall particles. For these particles the self image electrostatic interaction diverges, leading to an infinite constant which must be  renormalized.
We use $1 \times 10^6$ attempted particle moves to equilibrate and $100$ moves per particle to create a new state. After $40000$ states, we calculate the average ionic density profiles.

\section{Modified Poisson-Boltzmann Equation}\label{mpbe}

To test the simulation method developed above, we compare the results with a modified Poisson-Boltzmann~(mPB) equation.   The mPB equation is constructed to account approximately for
the ion-image and charge-charge correlations near an interface and is expected to work well in the weak coupling limit.   
It was tested against MC simulation for a single dielectric interface
with 1:1 electrolyte and was found to be very accurate. 
Therefore, we expect that for a slab geometry, a suitably modified mPB equation will also remain very accurate, allowing us to test the new simulation method.   

The electrostatic potential between two negatively charged dielectric surfaces satisfies the exact Poisson equation
\begin{equation}\label{pb}
\nabla^2\phi(z)=-\frac{4 \pi}{\epsilon_w}\left[Q \rho_+(z) - Q \rho_-(z)  \right] \ ,
\end{equation}
where  $\phi(z)$ is the mean electrostatic potential at a distance $z$ from the first plate (for simplicity it is placed at $z=0$), $\rho_+(z)$ and $\rho_-(z)$ are the concentrations of cations and anions derived from salt and surface dissociation.  In equilibrium, all the ions will be distributed
in accordance with the Boltzmann distribution, $\exp(- \beta \omega_s)$ , where $\omega_s$ is the potential 
of mean force of ion of type $s$. We will approximate $\omega_s$ by the mean-electrostatic potential
plus the correlation contribution, $W(z)$.  
Suppose that there are $N_+$ cations and counterions, and $N_-$ anions per square Angstrom.  Their
distributions are then given by,
\begin{equation}\label{rho+}
\rho_+(z)=N_+\frac{e^{-\beta Q\phi(z)-\beta W(z)}}{\int_{r_c}^{L-r_c} dz\ e^{-\beta Q\phi(z)-\beta W(z)}} \ ,
\end{equation}
\begin{equation}\label{rho-}
\rho_-(z)=N_-\frac{e^{\beta Q\phi(z)-\beta W(z)}}{\int_{r_c}^{L-r_c} dz\ e^{\beta Q\phi(z)-\beta W(z)}} \ .
\end{equation}
If the expression for $W(z)$ is known, we can solve the mPB equation numerically to obtain the ionic density profiles. 

\begin{figure}[h]
\begin{center}
\includegraphics[width=7cm]{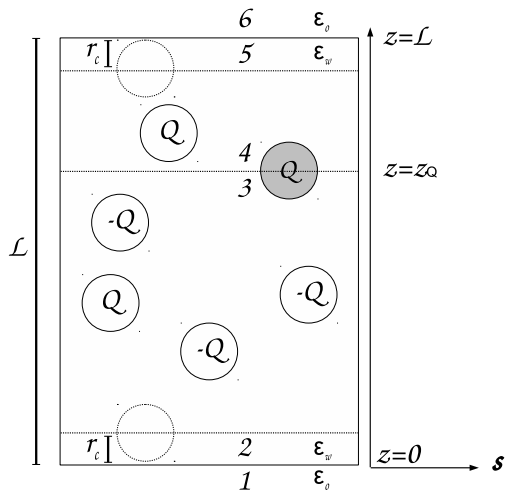}
\end{center}
\caption{Representation of an electrolyte in the region $r_c<z<L-r_c$ used to calculate $W(z)$.}
\label{fig2}
\end{figure}

The correlational and charge-image contribution $W(z)$ can be calculated approximately as follows. Consider $N_+$ ions (per \AA$^2$), with charge $Q$ and $N_-$ ions  (per \AA$^2$) with charge $-Q$, all with hydration radii $r_c$, confined between two neutral dielectric walls with separation $L$, see Fig.~\ref{fig2}. 
Due to the hardcore repulsion of ions from the surfaces, they are restricted to the region $z>r_c$ and $z<L-r_c$.
To keep the charge neutrality of this region,  we introduce a uniform neutralizing background charge density $-Q N_c/(L-2 r_c)$, where $N_c=N_+-N_-$.  In the exterior regions $z<0$ and $z>L$ the dielectric constant is $\epsilon_o$, while in the interior region it is $\epsilon_w$.  The function $W(z)$ then corresponds, approximately, to the energy penalty that an ion located at a distance $z$ from one of the surfaces feels due to asymmetry of its ionic atmosphere and due to its hard core repulsion from the wall. To obtain this potential, we calculate the Green's function for a system of differential equations: Laplace equation,
\begin{equation}
\nabla^2 \phi({\bf s},z)=0 \ ,
\end{equation}
in the region with no electrolyte and a linearized Poisson-Boltzmann~(LPB) equation,
\begin{equation}
\nabla^2 \phi({\bf s},z)=\kappa^2 \phi({\bf s},z) \ ,
\end{equation}
in the region accessible to ions~\cite{LeFl01}, where $\phi({\bf s},z)$ is the potential at position $({\bf s},z)$ in cylindrical coordinates and $\kappa=\sqrt{4\pi \lambda_{B}(N_+ + N_-)/L}$ is the inverse Debye length. LPB equation is used to account for the electrostatic correlations between the ions~\cite{Le02}.  
For 1:1 electrolyte, linearization of the Poisson-Boltzmann equation is justified since the ionic interactions are weak.   Because of the azimuthal symmetry of the problem, it is convenient to work with the Fourier transform of the potential, $\hat\phi({\bf k},z)$ defined in terms of~\cite{LeFl01},
\begin{eqnarray}
\phi({\bf s},z)=\dfrac{1}{2\pi}\int_{0}^{\infty}dk\ k J_0(ks)\hat\phi({\bf k},z) \ .
\end{eqnarray}
For dielectric interfaces between hydrocarbons and water,  $\epsilon_o/\epsilon_w \ll 1$, so that to leading order
we can set $\epsilon_o=0$. Taking into account the continuity of the electrostatic potential and of the normal component of the displacement field at $z=0$, $z=r_c$, $z=z_Q$, $z=L-r_c$ and $z=L$,  the Fourier transform of the electrostatic potential in the region 3, see Fig.~\ref{fig2}, can be calculated to be:
\begin{eqnarray}
\hat\phi_3({\bf k},z)=A_3 e^{pz}+B_3 e^{-pz} \ ,
\end{eqnarray}
where ${\bf k}$ is the wave vector and $p=\sqrt{\kappa^2 + k^2}$. The constants are given by:
\begin{eqnarray}
\begin{array}{l}

A_3= \frac{\pi Q f_1}{\epsilon_w f_3}\left(e^{p z_Q} \left[ (p+k) e^{ 2r_cp-r_ck}+(p-k) e^{2r_cp+r_ck}\right] \right.+\nonumber \\
\left. e^{-p z_Q}\left[(p+k)  e^{2 L p+r_c k}+(p-k) e^{2 L p-r_c k}\right] \right) \ , \nonumber \\

B_3=A_3 e^{2 r_c p} \frac{f_2}{f_1} \ , \nonumber \\

f_1=p \cosh (k r_c)+k \sinh (k r_c) \ , \nonumber \\

f_2=p \cosh (k r_c)-k \sinh (k r_c) \ , \nonumber \\

f_3=p \left(e^{2 L p} f_1^2 - e^{4 r_c p} f_2^2\right) \ . \nonumber

\end{array}
\end{eqnarray}

Considering $s \rightarrow 0$ and $z \rightarrow z_Q$, we find (now ommiting subscript 3):
\begin{eqnarray}
\phi(z_Q)= \dfrac{Q}{2 \epsilon_w}\int_{0}^{\infty}dk\ k\ (\ \frac{f_1}{f_3}e^{2 p z_Q}
\left[ (p+k) e^{ 2r_cp-r_ck}\right.+\nonumber \\
\left. (p-k) e^{2r_cp+r_ck}\right]  + \frac{f_2}{f_3}e^{- 2 p z_Q + 2 r_c p }\left[(p+k)  e^{2 L p+r_c k}\right.+\nonumber \\
\left. (p-k) e^{2 L p-r_c k}\right] ) + func \ . \nonumber \\
\end{eqnarray}
where $func$ does not depend on the ion-plate distance $z_Q$ and can be ignored. The work necessary to insert an ion at position $z=z_Q$ from the interface,  
can be calculated using the G\"untelberg~\cite{Gu26} charging process:
\begin{eqnarray}\label{image_potential_ii}
W(z_Q)= \dfrac{Q^2}{4 \epsilon_w}\int_{0}^{\infty}dk\ k\ ( \ \frac{f_1}{f_3}e^{2 p z_Q}
\left[ (p+k) e^{ 2r_cp-r_ck}\right.+\nonumber \\
\left. (p-k) e^{2r_cp+r_ck}\right]  + \frac{f_2}{f_3}e^{- 2 p z_Q + 2 r_c p }\left[(p+k)  e^{2 L p+r_c k}\right. +\nonumber \\
\left. (p-k) e^{2 L p-r_c k}\right] ) \ . \nonumber \\
\end{eqnarray}
The interaction potential $W(z)$ can now be used in the mPB equation, Eq.~\eqref{pb}, to account for the
charge-image interaction and the polarization of the ionic atmosphere.

\section{Results}\label{r}

\begin{figure}[t]
\begin{center}
\includegraphics[width=8.5cm]{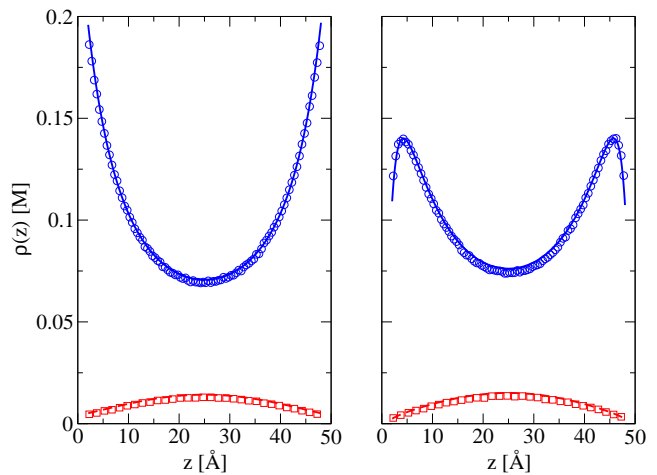}
\end{center}
\caption{Ionic density profiles for the cases: $\epsilon_o=\epsilon_w=80$, left, and $\epsilon_o=2$ and $\epsilon_w=80$, right. The plate charge density is $-0.02~$C/m$^2$, the distance between plates is $L=50~$\AA\ and the monovalent salt concentration is $10~$mM. Symbols represent the simulation data and lines represent the solution of the mPB equation, Eq.~\eqref{pb}. Solid lines and circles represent cations, while dashed lines and squares represent the anions.}
\label{fig3}
\end{figure}
We first show the comparison between the ionic concentrations obtained using mPB Eq.~\eqref{pb} and the results of MC simulations. In Fig.~\ref{fig3} we show the ionic density profiles between two interfaces for the cases: (1) where $\epsilon_o=2$ and $\epsilon_w=80$, for which $W(z)$ is approximately given by Eq.~\eqref{image_potential_ii}, and (2) where $\epsilon_o=\epsilon_w=80$, for which $W(z)$ is zero. The agreement between theory and simulations is very good, validating both methods.

\begin{figure}[h]
\begin{center}
\includegraphics[width=8.5cm]{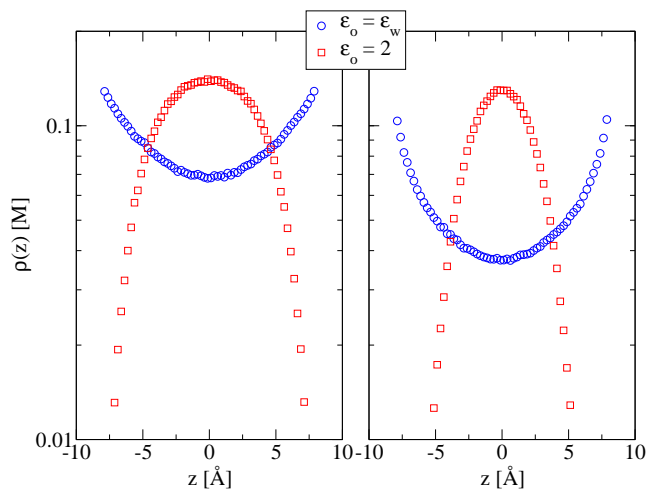}
\end{center}
\caption{Couterion density profile for the cases $\alpha_j=3$, left, and $\alpha_j=5$, right. The plate charge density is $-0.02~$C/m$^2$, the distance between plates is $L=20~$\AA.}
\label{fig4}
\end{figure}
\begin{figure}[h]
\begin{center}
\includegraphics[width=8.5cm]{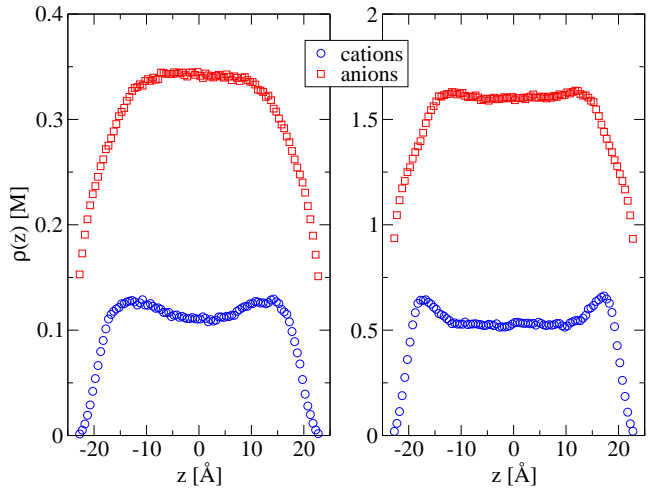}
\end{center}
\caption{Ionic density profiles for  $\epsilon_o=2$ and 3:1 salt confined between neutral surfaces, for various salt concentrations. On the left, $100~$mM of salt, while on the right, $500~$mM. The distance between plates is $L=50~$\AA. Circles represent trivalent 
cations and squares monovalent anions.}
\label{fig5}
\end{figure}

We next explore the effects of the dielectric heterogeneity for strongly correlated systems. The valency of the counterions is modified to $\alpha_j=3$ and $\alpha_j=5$. The other parameters are kept the same as before. 
If $\epsilon_w=\epsilon_o$, most of the multivalent ions adsorb to the charged wall forming a strongly correlated quasi 2D one  component plasma, see Fig.~\ref{fig4}. However, in the case of large dielectric contrast between solvent and the confining surface, the counterions experience a strong charge-image repulsion from the surface, see Fig.~\ref{fig4}. This can significantly affect the interaction between charged dielectric  bodies inside an electrolyte solution~\cite{WaWa13}. Finally, in Fig.~\ref{fig5}, we show the ionic distribution for 3:1 electrolyte between
neutral dielectric surfaces.  

\section{Conclusions}\label{c}

In this paper we developed a new simulation approach to study electrolytes and ionic liquids in a dielectric slab geometry. The method is easy to implement and is comparable in time consumption with the regular 3D Ewald summation method. In the weak coupling limit we also presented a mean-field theory which allows us to predict the ionic distribution between the dielectric charged plates. 
The agreement between both approaches, theoretical and simulational, is very good, validating both methods. Examples of ionic density profiles for strongly correlated systems are also presented. Finally, the simulation method developed here can be used to explore the interactions between colloidal particles with strong dielectric contrast.  This will be the subject of the future work.

\section{Acknowledgments}
This work was partially supported by the CNPq, INCT-FCx, and by the US-AFOSR under the grant 
FA9550-12-1-0438.

%

\end{document}